\documentclass{article}

\usepackage{arxiv}

\usepackage[utf8]{inputenc} 
\usepackage[T1]{fontenc}    
\usepackage{hyperref}       
\usepackage{url}            
\usepackage{booktabs}       
\usepackage{amsfonts}       
\usepackage{nicefrac}       
\usepackage{microtype}      
\usepackage{lipsum}		
\usepackage{graphicx}
\usepackage{natbib}
\usepackage{doi}
\usepackage{fancyhdr,lastpage,url,graphicx}
\usepackage{wrapfig}
\usepackage{color}
\usepackage{comment}
\usepackage{soul}
\usepackage{setspace} 
\usepackage[caption=false]{subfig}
\usepackage{wrapfig}
\usepackage{color}
\usepackage{cite}
\usepackage{units}
\usepackage{bmpsize}
\usepackage{array}
\usepackage{graphicx}
\usepackage{float}
\usepackage{amsmath}
\usepackage{amssymb}
\usepackage{balance}
\usepackage{empheq}
\usepackage[many]{tcolorbox}
\usepackage{algorithm}
\usepackage{algpseudocode}

\DeclareMathOperator*{\argmin}{argmin}   

\newtheorem{thm}{Theorem}[section]

\newtheorem{rem}[thm]{Remark}

\title{Control Co-design of Actively Controlled Lightweight Structures for High-acceleration Precision Motion Systems}

\author{ Jingjie~Wu  \\
	Walker Department of Mechanical Engineering\\
	The University of Texas at Austin\\
	Austin, TX, 78712 \\
	\texttt{wujingjie@utexas.edu} \\
	\And
	{Lei~Zhou} \\
	Walker Department of Mechanical Engineering\\
	The University of Texas at Austin\\
	Austin, TX, 78712 \\
	\texttt{Lzhou@utexas.edu} \\
}

\date{}

\renewcommand{\shorttitle}{\textit{arXiv} Template}

\begin{document}
\maketitle

\begin{abstract}
Precision motion stages are an essential part of a wide range of manufacturing equipment, and their motion performance are critical to the quality and throughput of the systems. The drastically increasing demand for higher manufacturing throughput in various processes necessities the development of next-generation motion systems with reduced moving weight and high control bandwidth. However, the reduction of moving stage's weight can lower the stage's structural resonance frequencies, making the hardware dynamics and controller design problem strongly coupled. Aiming at this challenge, this paper proposes a new formulation of nested hardware and control  co-design framework for precision motion stages. The proposed framework explicitly optimizes the closed-loop control bandwidth with guaranteed robustness, and explicitly considers the limits in the physical system.  Two case studies, including a motivating example using lumped-parameter mechanical system  and a finite-element-simulated lightweight motion stage, are being used to evaluate the effectiveness of the proposed nested CCD framework. Simulation results show that the proposed nested CCD framework has 42\% of weight reduction and 28\% bandwidth improvement compared with a sequential design baseline, which demonstrates the efficacy of the proposed approach.
\end{abstract}


\section{Introduction}\label{sec:intro}

Precision positioning stages are an essential part of a wide range of manufacturing machines including machine tools, 3D printers, and wafer scanners, and the stage's  positioning accuracy and speed are critical to the manufacturing quality and productivity of the associated process. In recent years, the demand for higher throughput and reduced energy consumption in manufacturing equipment necessitates the development of next-generation precision motion systems with light moving weight \citep{oomen2013connecting}. However, as the stage's weight reduces, its structural resonance frequencies will decrease to near or even within the control bandwidth, which leads to strong coupling between the system's hardware design decisions and its controller design. 
This fact motivates the study for new design approaches to better exploit the synergy in the hardware and controller designs, thereby enabling new motion stages with improved overall performance. 
This fact motivates the study for new design approaches to better exploit the synergy in the hardware and controller designs, thereby enabling new motion stages with improved overall performance.

In the past decade, combined hardware and control co-design also referred as control co-design (CCD), is receiving drastically increasing research attention as a means to address the aforesaid challenge of hardware-control design coupling \citep{allison2013multidisciplinary, allison2014co,  nash2019combined, deshmukh2016multidisciplinary}. CCD is a dynamic system design methodology where the system's hardware design parameters and control policy are being optimized in a uniform framework \citep{garcia2019control}. Compared to conventional sequential design where the physical plant is designed first followed by controller design, the CCD approach has enhanced design flexibility and explores a larger feasible design space, and thus enables design solutions with improved closed-loop performance \citep{fathy2001coupling}. Prior studied CCD algorithms can be categorized into four groups: sequential, iterative, nested, and simultaneous \citep{sundarrajan2021towards}. The sequential strategy decouples the hardware and control optimization problems and thus simplifies the solution; the outcome is however typically sub-optimal. On the other extreme, the simultaneous algorithms fully couple the hardware and control optimizations leading to guaranteed overall  optimally \citep{allison2014co}, however they are typically computationally expansive  and are often intractable or impractical for large-size problems \citep{sundarrajan2021towards}. The iterative and nested approaches are between the two extremes and provide a balance between computational complexity and performance. While the iterative algorithms typically cannot guarantee the system-level optimality, the nested approach can, under certain conditions~\citep{herber2019nested}.

It has been long recognized that the controller design for precision motion stages should be accounted for during the hardware design phase \citep{rankers1998machine}, and several research initiatives have focused on creating effective design approaches for systems with hardware-control design coupling. In the 1980s and 1990s, a number of studies on control-structure interaction (CSI) have been carried out focusing on simultaneously reducing structure's weight while meeting requirements on dynamics \citep{hale1985optimal,cheng1992control}. While these  works obtained promising results, they largely focused on specific tools and systems, for example truss structures for aerospace applications. In the 2000s, systematic CCD theory and methods have been created \citep{allison2013multidisciplinary} and have shown promise for applications including thermal-fluid systems \citep{nash2019combined}, wind turbines\citep{deshmukh2016multidisciplinary}, and automobile components \citep{allison2014co}. However, these formulations typically require an analytical system model for parameter optimization, which is difficult to obtain for motion stages due to their geometric complexity. 
In recent years, driven by the demand for higher throughput in the photolithography process, van der Veen et al.~\citep{van2017integrating} and Wang et al.~\citep{wang2019integrated} studied the integration of control and topology optimization for motion stages with 2D simple structures. Ding et al.~\citep{ding2020optimal} explored the use of genetic algorithm to optimize actuator/sensor positions, and Wang et al.~\citep{wang2019simultaneous} studied the use of simultaneous CCD strategy to minimize stage vibration energy. 
While these efforts demonstrated the initial study of adapting CCD approaches for precision motion systems, frequency-domain control performance specifications (e.g. control bandwidth, maximum disturbance sensitivity, etc) are often not explicitly considered in the problem formulations, which are of critical importance in motion systems. In addition, physical system bounds, such as actuator's capability that bounds control effort signal, are typically not considered. These facts limit the  applicability of the CCD approaches for practical precision motion systems.

In order to address the aforementioned challenges, in this work, we propose a CCD framework for precision motion stages with the following key features: (a) explicitly optimizes for the closed-loop control bandwidth, (b) incorporating a constraint on the maximum disturbance sensitivity to guarantee control robustness, and (c) explicitly considering bounds for control effort signals. 
The proposed CCD framework takes a nested formulation to ensure optimality, where the outer loop optimizes the hardware parameters, and the inner loop synthesizes a controller optimizing for the control bandwidth while satisfying robustness constraints.
The proposed approach is evaluated on two case study systems, including a motivating example of a lumped mass-spring-damper system and a rib-reinforced lightweight motion stage. Simulation results show that the proposed approach has 42\% of weight reduction and 28\% bandwidth improvement compared with the baseline sequential design approach, which shows promise as an effective design tool for lightweight motion stages. 

The rest of this paper is organized as follows. Section~\ref{sec:statement} presents the problem statement. Section~\ref{sec:CCD} presents the proposed nested CCD framework and algorithm for precision motion systems. Section~\ref{sec:simulations} shows the simulation evaluation for the proposed CCD framework. Conclusion and suggested future work are discussed in Section~\ref{sec:conclusion}.

\section{Problem Statement}\label{sec:statement}

The dynamics of a motion stage considering its flexible dynamics can be written as
\begin{align} 
M(\theta)\ddot{x}+D(\theta)\dot{x}+K(\theta)x &= B(\theta)u,\label{eqn:mech_EOM_1}\\
    y &= C(\theta)x, \label{eqn:mech_EOM_2}
\end{align}  
where $x$ is the state variable including both the stage's rigid-body displacements and its flexible modal displacements, $M$, $D$, $K$ are the mass, damping, and stiffness matrices,  respectively, $B$ is the input matrix, $C$ is the measurement matrix, $u$ is the control input, $y$ is a vector of measurement signals, and $\theta$ is a vector of hardware design parameters.

The CCD problem for system \eqref{eqn:mech_EOM_1} and \eqref{eqn:mech_EOM_2} can be roughly formulated as: find a feasible set of  hardware parameter selection $\theta$ and feedback controller design that can simultaneously maximize the closed-loop control bandwidth and minimize the moving stage's weight while satisfying robustness criteria.

\section{Nested Control Co-Design Algorithm}\label{sec:CCD}

\begin{figure}[t!]
\centering
\subfloat{
\includegraphics[trim={15mm 9mm 15mm 15mm},clip, width =0.5\columnwidth, keepaspectratio=true]{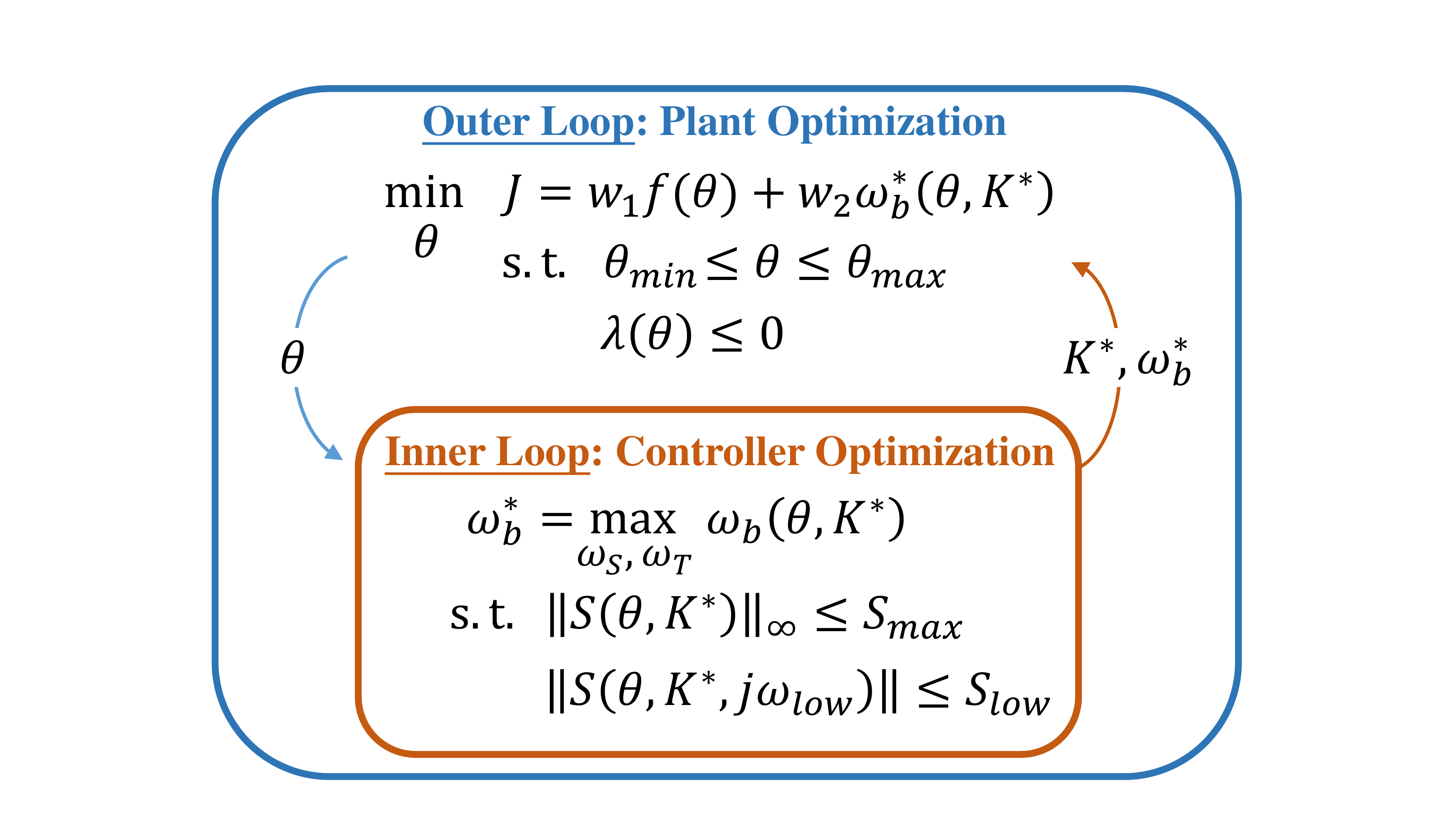}}
\vspace{-2mm}
\caption{Overview diagram of proposed nested CCD formulation for lightweight motion system.} 
    \vspace{-6mm}
\label{fig:nest_algorithm}
\end{figure}

This section introduces the CCD problem formulation and a proposed algorithm for the CCD problem for lightweight motion systems introduced in Section~\ref{sec:statement}. As discussed in Section~\ref{sec:intro}, there exist four different formulations for CCD frameworks including sequential, iterative, nested, and simultaneous. Among these CCD strategies, the nested and simultaneous algorithms can provide guaranteed optimality, and the nested formulation typically allows a reduced computational load to solve comparing with the simultaneous CCD formulation. 
The nested CCD framework uses a two-level optimization problem to conduct the system design, where the outer loop optimizes plant variables and the inner loop synthesizes an optimal controller for the plant design determined by the outer loop \citep{herber2019nested}. By partitioning the overall system optimization into two subproblems with reduced size and complexity, the nested CCD formulation requires reduced computation than simultaneous CCD formulations. In addition, the inner loop in a nested CCD formulation can utilize existing tailored algorithms for optimal control problems  such as robust MPC \citep{9483216}, and guaranteed cost control \citep{zeng2019integrated}, which provides the benefit of efficiency improvement for the overall CCD problem solving \citep{sundarrajan2021towards}.

Due to the aforesaid advantages of the nested CCD algorithms, in this paper,  a nested CCD formulation is selected for our problem. Fig.~\ref{fig:nest_algorithm} shows an overview of the proposed nested CCD formulation targeting lightweight motion systems. Here, the overall system objective function $J$  is defined as a weighted sum of the hardware cost $f(\theta)$ and the closed-loop bandwidth $\omega_b^*$. 
The proposed formulation includes two loops: the outer loop solves the hardware parameter that optimizes the overall objective function, and the inner loop solves an optimal control problem that optimizes for the control bandwidth while satisfying robustness criteria. Note that in the design optimization for  motion stages, the weight of the motion stage is selected as the hardware system cost; however, this objective function can be selected otherwise to reflect other design considerations. The rest of this section introduces the inner and outer loops.

\subsection{Inner Loop Optimization}

This section introduces the formulation for the inner loop in the nested CCD problem in Fig.~\ref{fig:nest_algorithm}. To effectively design a controller that optimizes for frequency-domain design specifications, we selected the  mixed sensitivity $H_{\infty}$ robust control algorithm as a building block. To make this paper self-contained, we first briefly introduce  the general formulation of the mixed sensitivity $H_{\infty}$ control. After that, we discuss the specific inner loop algorithm in the proposed nested CCD formulation.

\subsubsection{Mixed Sensitivity $H_{\infty}$ Robust Control}\label{sec:mixsym}

\begin{figure}[t!]
\centering
\subfloat{
\includegraphics[trim={70mm 58mm 90mm 55mm},clip,width = 0.5\columnwidth, keepaspectratio=true]{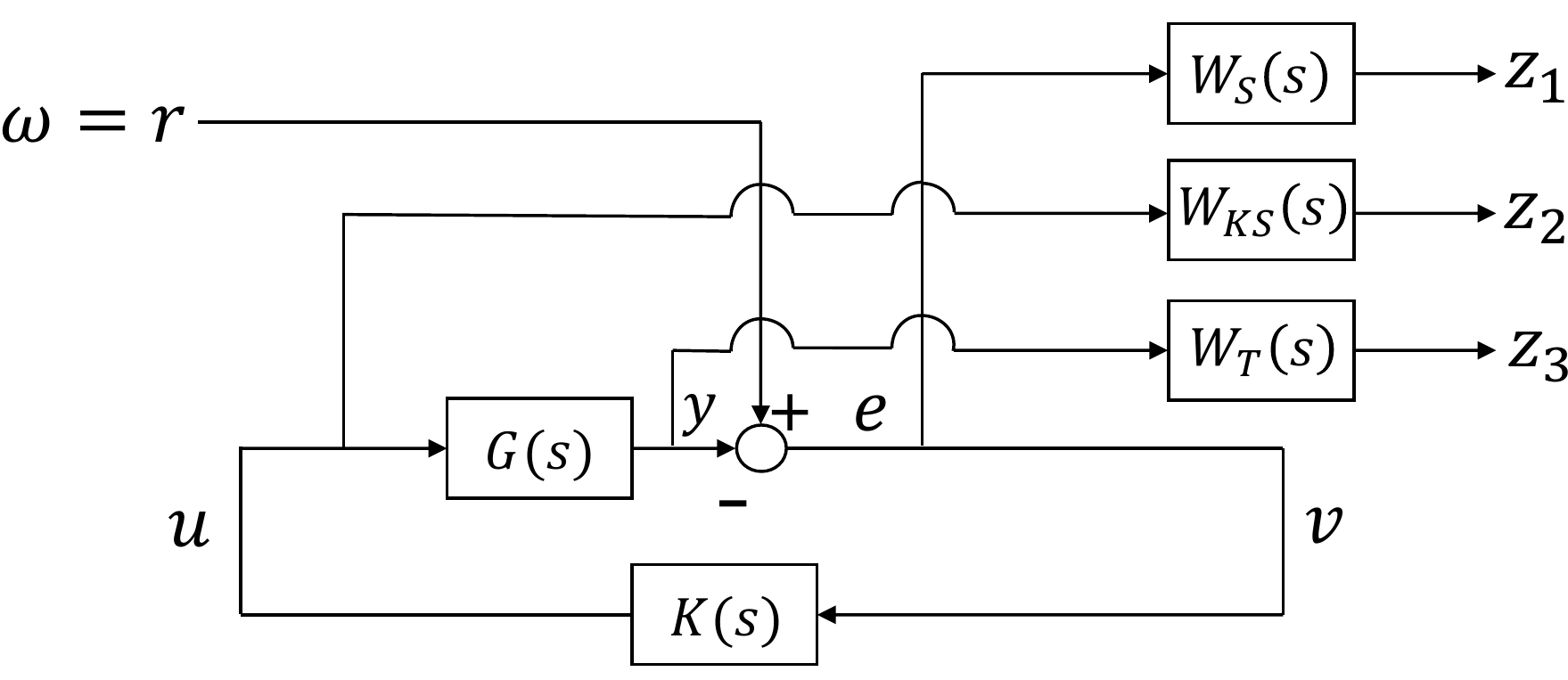}}
\vspace{-2mm}
\caption{Mixed sensitivity $H_\infty$ control design framework with plant model $G$,  controller $K$, reference $r$, control input $u$, error $e$, performance output $z_1, z_2, z_3$, design weighting filters $W_S, W_{KS}, W_T$.} 
\vspace{-6mm}
\label{fig:mixed_syn}
\end{figure}

Mixed sensitivity $H_{\infty}$ control design  is a robustness control design technique that can simultaneously shape the frequency responses of the dynamic system and balance the trade-off between robustness and performance \citep{skogestad2007multivariable, mashrafi2020optimal}. Fig.~\ref{fig:mixed_syn} shows a diagram for a closed-loop system and illustrates the mixed sensitivity $H_{\infty}$ design framework. Here, $G(s)$ is the plant, $K(s)$ is the controller, $u$ is the control input signal, $y$ is the measurement signal, $e$ is the tracking error, $r$ is the reference, $W_S$, $W_{KS}$, and $W_T$ are weighting filters. There are three performance output signals: 
$z_1=W_{S}e$ is the weighted tracking error signal that reflects the tracking bandwidth and disturbance rejection objectives, $z_2 = W_{KS}u$ is the weighted control input signal which imposes bounds to the control efforts, and $z_3 = W_{T}y$ is the weighted measurement signal that characterizes  the noise attenuation and robustness objective. The optimal mixed sensitivity $H_{\infty}$ controller $K^{*}$ is synthesized via the following optimization problem:
\begin{align}  \label{eqn:mix_syn}
    K^* = \argmin_K \Bigg\| \begin{bmatrix}  W_{S}S \\ W_{KS}KS\\W_{T}T  \end{bmatrix} \Bigg\|_{\infty},
\end{align}
where $S = (I+GK)^{-1}$, $KS = K(I+GK)^{-1}$, and $T = GK(I+GK)^{-1}$ are the sensitivity, control sensitivity, and complementary sensitivity functions.

In the mixed sensitivity $H_{\infty}$ design, the selection of weighting filters parameters is critical for the control performance. The systematic selection of filter parameters is discussed in \citep{ortega2004systematic}, which is briefly summarized here. For a multi-input-multi-output (MIMO) system,  the weighting filters take the form of diagonal matrices of single-input-single-output filters as

\vspace{-4mm}
\begin{small}
\begin{align}
    W_S(s)&=\mathrm{diag}\{W_1(s),\cdots,W_1(s)\},\\ W_{KS}(s)&=\mathrm{diag}\{W_2(s),\cdots,W_2(s)\},\\ W_{T}(s)&=\mathrm{diag}\{W_3(s),\cdots,W_3(s)\},
\end{align}
\end{small}
where
\begin{small}
\begin{align}\label{eqn:mix_filter}
 W_1(s)&=\frac{s/M_S+\omega_S}{s+\omega_S A_S},
\\
 W_2(s) &= \frac{c_K(s+\omega_K)}{M_K(s+c_K\omega_K)},
\\
 W_3(s) &= \frac{s+(1/A_l)\omega_T}{A_u s+\omega_T}.
\end{align}
\end{small}

Here, $\theta_W = [M_S, \omega_S, c_K, \omega_K, M_K, A_l, \omega_T, A_u]^\top$ is a vector for the weighting filter parameters. The selection rule for these parameter values is summarized in Table~\ref{tab:parameter_rule}.

\begin{table}[t!]
\renewcommand{\arraystretch}{1.0}
  \caption{Weighting filter parameters selection \citep{ortega2004systematic}.}\label{tab:parameter_rule}\renewcommand\baselinestretch{0.8}\selectfont
\vspace{-2mm}
\begin{center}\begin{small}
  \begin{tabular}{ |m{0.7cm} |p{6.3cm} |}
\hline
Value & Selection Rule \\
\hline
    $M_S$ & Upper bound of maximum peak of sensitivity function $S$ \\ \hline
    $A_S$ & Limit steady-state tracking error at the low frequency range \\ \hline
    $M_K$ & Bound of control effort signal\\ \hline
    $\omega_K$ & A frequency up to which the control effort bound is imposed \\ \hline
    $c_K$ & Determines robust stability against additive uncertainty at the high frequency range; typical value between $10^3$ and $10^5$.  \\ \hline 
    $A_u$ & Determines controller roll-off at high frequency and improve the robust stability against multiplicative uncertainty and noise attenuation; typical value between $10^{-2}$ and $10^{-4}$.  \\ \hline  
\end{tabular}
\end{small}
\end{center}
    \vspace{-6mm}
\end{table}

\subsubsection{$H_{\infty}$-Based Max-Bandwidth Robust Control}  \label{sec:filter_para}

This section discusses the specific formulation for the inner loop in our nested CCD formulation. To enable high-performance motion control in precision positioning stage application, the inner loop of the CCD problem solves for a  controller design that maximizes the control bandwidth $\omega_b$ of the closed-loop system while maintaining sufficient robustness and tracking error criteria  as
\begin{align}  \label{eqn:inner}
\begin{split}
    \max_{\theta_w} ~~~&\omega_b(\theta,K^*),
\\
    \mathrm{s.t}.~~~ &\| S(\theta,K^*) \|_{\infty} \leq S_{max},
\\
   & \|  S(\theta,K^*,j\omega_{low})\| \leq S_{low}.
\end{split}
\end{align}
Here, $S_{max}$ is the bound for the maximum value of the sensitivity function, $S_{low}$ is the bound for sensitivity singular value in low-frequency range  evaluated at a specified frequency $\omega_{low}$. 
Note that the control bandwidth $\omega_b$  in this work is defined as the frequency at which the maximum singular value of $S$  reaches $-3~\mathrm{dB}$ for the first time, i.e.~$\|S(\theta,\omega_b,K)\|_2 = -3~\mathrm{dB}$. This definition follows that in reference \citep{skogestad2007multivariable}.

The mixed sensitivity $H_\infty$ control algorithm is selected to determine an optimal controller $K^*$ by solving \eqref{eqn:mix_syn} with the plant and weighting filter parameters determined. With the plant parameter, $\theta$ fixed in the inner loop optimization, the optimal controller $K^*$ is uniquely determined by the weighting filter parameters $\theta_w$. Therefore the decision variables for the inner-loop optimization problem \eqref{eqn:inner} is $\theta_w$. 

One challenge in the solving of the inner loop problem \eqref{eqn:inner} is that there does not exist a closed-form expression for the gradient of the objective and constraints. As a result, optimization algorithms based on the direct search approach is needed to find the optimal weighting filter parameters $\theta_w$, and the computational cost for such algorithm increases significantly with respect to the problem size. To reduce the computational load, we conducted a simplification to the inner loop optimization problem by selecting a subset of $\theta_w$ as the decision variables. Specifically, parameters $\omega_S$ and $\omega_T$ are selected as the decision variables in \eqref{eqn:inner}, which reduces the inner loop problem to 
\begin{align}  \label{eqn:inner_reduced}
\begin{split}
    \max_{\omega_S,\omega_T} ~~~&\omega_b(\theta,K^*),
\\
    \mathrm{s.t.}~~~ &\| S(\theta,K^*) \|_{\infty} \leq S_{max},
\\
   & \|  S(\theta,K^*,j\omega_{low})\| \leq S_{low}.
\end{split}
\end{align}
\begin{rem}

Our decision variables in \eqref{eqn:inner_reduced} are selected as  $\omega_S$ and $\omega_T$ because they are the break frequencies of the weighting filters determining $S$ and $T$; therefore the control bandwidth $\omega_b$ is highly sensitive to these two variables. 
The values of other parameters in $\theta_w$ are selected based on the physical system bounds (e.g. maximum allowed control effort signal) and desired control specification (e.g. maximum allowed singular value of sensitivity). The specific parameter selections for  detailed problems are discussed in Section~\ref{sec:simulations}.
\end{rem}

The algorithm that we used to solve problem~\eqref{eqn:inner_reduced} is discussed as follows. For each given plant parameter $\theta$, we conduct a parameter sweep for $\omega_S$ and $\omega_T$, and a mixed sensitivity $H_{\infty}$ controller $K^*$ is synthesized for every $(\omega_S, \omega_T)$ in the searching range. Then the optimal control bandwidth  $\omega_b^*$ is found among the feasible controller designs. 
Note that the initial searching ranges for $\omega_S$ and $\omega_T$ are chosen to be large to capture the optimal solution satisfying the constraints for the initial plant parameters. Starting from the second iteration, the searching range can be chosen to be a small range around the optimal values of $\omega_S$ and $\omega_T$ from the previous iteration. This approach is valid since every step in the outer loop the plant parameters $\theta$ is updated with a small step size and can effectively reduce the total computational effort for the parameter sweep.

\subsection{Outer loop optimization}  \label{sec:outer_loop}
The outer loop in Fig.~\ref{fig:nest_algorithm} searches for the plant parameter $\theta$ that optimizes the overall objective as 
\begin{align}   \label{eqn:outer}
\begin{split}
    \min_{\theta} ~~~&J = w_1f(\theta)+w_2\omega_b^*(\theta,K^*),
\\
    \mathrm{s.t.} ~~~~~~&\theta_{min} \leq \theta  \leq \theta_{max},
\\
    & \lambda(\theta) \leq 0,
\end{split}
\end{align}
where $\theta_{min}$ and $\theta_{max}$ are the bounds for plant parameter $\theta$, $\lambda(\theta)$ represents the additional constraints on $\theta$, $K^*$ and $\omega_b^*$ are the optimal controller design and maximum control bandwidth, respectively, which are computed in the inner loop and passed to the outer loop. Herein, the objective function $J$ contains the weighted sum of a specified objective function $f(\theta)$ and the optimal control bandwidth $\omega_b^*$ solved in the inner loop, and $w_1$ and $w_2$ are the weights for plant and control design objectives, respectively. 

\begin{rem}
To enable effective design optimization, the selected plant objective $f(\theta)$ should have a certain trade-off with the achievable  control bandwidth. In this work, $f(\theta)$ is selected to be the weight of the moving stage. Since reducing the stage's weight can lower the structure's resonance frequency and thus limits the achievable control bandwidth, a meaningful trade-off can be made by solving \eqref{eqn:outer}.
\end{rem}

The outer loop optimization problem \eqref{eqn:outer} is solved via the steepest-gradient descent algorithm. The computation of the gradients is introduced below.

The gradient of the objective $J$ with respect to $\theta$ is
\begin{equation}  \label{eqn:J_grad}
\frac{\partial J}{\partial \theta} = w_1\frac{\partial f(\theta)}{\partial \theta} + w_2\frac{\partial \omega^*_b (\theta,K^*)}{\partial \theta},
\end{equation}
where ${\partial f(\theta)}/{\partial \theta}$ is available analytically or numerically. The second term  ${\partial \omega_b}/{\partial \theta}$ can be found by applying the implicit function theorem \citep{krantz2012implicit} as
\begin{equation} \label{eqn:BW_theta}
\medmuskip=0mu
\thinmuskip=0mu
\thickmuskip=0mu
\frac{\partial \omega_b}{\partial \theta} = - \frac{\partial \|S(\theta,\omega_b,K)\|_2 }{\partial \theta} \big(\frac{\partial \|S(\theta,\omega_b,K)\|_2 }{\partial \omega_b}\big)^{-1},
\end{equation}
The values of ${\partial \|S\|_2 }/{\partial \theta}$ and ${\partial \|S\|_2 }/{\partial \omega_b}$ can be found by applying the gradient of singular value $\sigma_i$ as of a general matrix $A$ as \citep{mukhopadhyay1982application}
\begin{equation} \label{eqn:sigma_G_p}
\medmuskip=0mu
\thinmuskip=0mu
\thickmuskip=0mu
\frac{\partial \sigma_i}{\partial p} = \mathrm{Real}\big[u_i^* \frac{\partial A}{\partial p}v_i\big],
\end{equation}
where $u_i$ and $v_i$ are the $i$-th column of the unitary matrices $U$ and $V$ from the singular value decomposition, i.e., $A = U\Sigma V^*$. Applying \eqref{eqn:sigma_G_p} to \eqref{eqn:BW_theta}, we have
\begin{align} \label{eqn:S_sigma_theta}
\frac{\partial \|S\|_2 }{\partial \theta_i} &= \mathrm{Real}\big[u_1^* \frac{\partial S}{\partial \theta_i}v_1\big], \\
\frac{\partial \|S\|_2 }{\partial \omega_b}&= \mathrm{Real}\big[u_1^* \frac{\partial S}{\partial \omega_b}v_1\big],
\end{align}
where $u_1$ and $v_1$ are the first columns of $U$ and $V$ corresponding to the maximum singular value $\sigma_1$ of matrix $S$. 
Consider
$\frac{\partial U^{-1}}{\partial x} = - U^{-1} \frac{\partial U}{\partial x} U^{-1}$,
we have 
\begin{align} \label{eqn:S_theta}
\medmuskip=-4mu
\thinmuskip=-4mu
\thickmuskip=-4mu
\frac{\partial S}{\partial \theta} &= -(I+GK)^{-1} \frac{\partial G}{\partial \theta}K (I+GK)^{-1},\\
\frac{\partial S}{\partial \omega_b} &= -(I+GK)^{-1}\Big( \frac{\partial G}{\partial \omega_b}K + G \frac{\partial K}{\partial \omega_b} \Big)(I+GK)^{-1},
\end{align}
where ${\partial G}/{\partial \theta}$, ${\partial G}/{\partial \omega_b}$, and ${\partial K}/{\partial \omega_b}$ can be found from the plant  dynamic model. With these aforesaid relationships, the gradient of the outer-loop objective function can be computed. Finally, combining \eqref{eqn:inner} and \eqref{eqn:outer}, the  nested CCD algorithm in Fig.~\ref{fig:nest_algorithm} can be solved.

\section{Simulation Evaluation}\label{sec:simulations}

Two case studies are used to evaluate the performance of the proposed nested CCD algorithm. Case~study~\#1 considers  a lumped-parameter mechanical system consists of two moving masses connected by a spring and a damper, where the spring stiffness is a function of the masses to mimic structural resonance of moving stages. This system is being used to study the impact of flexible dynamics on control performance, and how CCD methodology be used to better balance the trade-offs.  Case~study~\#2 evaluates the position control for a moving stage with rib-reinforced structure. 

\subsection{Case~Study~\#1: Lumped Mass-Spring Model}

\begin{figure}[t!]
\centering
\subfloat{
\includegraphics[trim={80mm 90mm 150mm 45mm},clip,width =0.3\columnwidth, keepaspectratio=true]{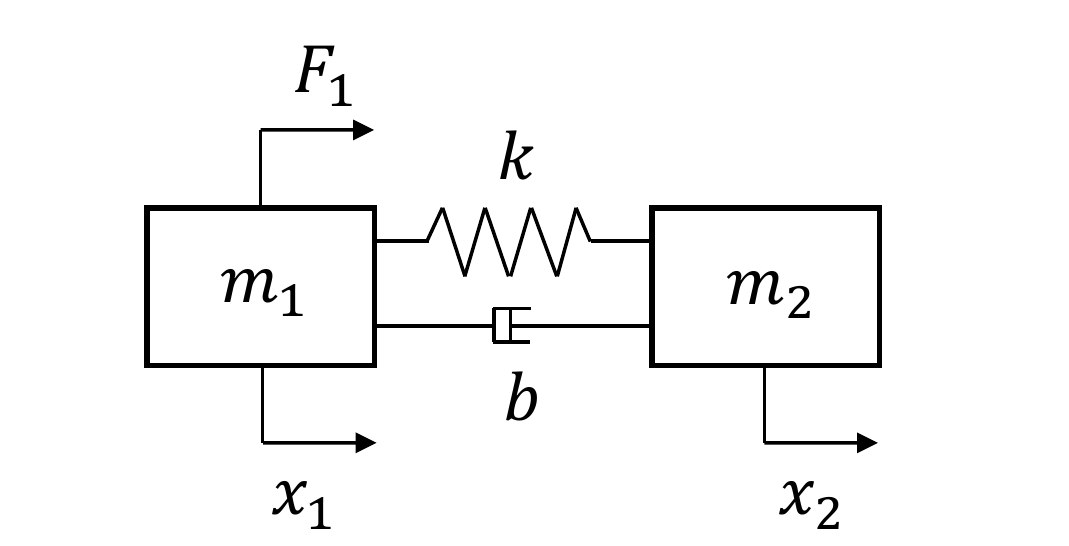}}
\vspace{-2mm}
\caption{Diagram of the two-mass-spring-damping system considered in Case study \#1. Here, the spring stiffness is selected as $k = 2m_1^4 +2m_2^4$ to mimic the structural resonance frequency trend under mass variation. Damping ratio is maintained at $\zeta = 0.01$.}
\vspace{-4mm}
\label{fig:benchmark}
\end{figure} 

Fig.~\ref{fig:benchmark} shows a diagram for the system being considered in case study \#1, which consists two moving masses $m_1$ and $m_2$ connected by a spring with stiffness $k$ and a damper with damping $b$. 
The control input is the force $F_1$, and the measurement is $x_1$ to mimic the collocated system dynamics \citep{rankers1998machine}. 
In order to mimic the property of continuum structures where the resonance frequency of flexible modes decreases as the structure's weight decreases, the stiffness $k$ is set to be a polynomial function of moving masses, and specifically, we selected $k = 2m_1^4 +2m_2^4$. 
The undamped dynamics of the system in Fig.~\ref{fig:benchmark} is
\begin{align}  \label{eqn:mass_spring}
& \begin{bmatrix} m_1 & 0 \\ 0 & m_2 \end{bmatrix}
\begin{bmatrix} \ddot{x_1} \\ \ddot{x_2} \end{bmatrix} 
+ \begin{bmatrix} k & -k \\ -k & k \end{bmatrix} \begin{bmatrix} x_1 \\ x_2 \end{bmatrix} 
= \begin{bmatrix}  1 \\ 0 \end{bmatrix} F_1, \\
&\qquad \qquad \qquad y = \begin{bmatrix} 1 & 0 \end{bmatrix} \begin{bmatrix}  x_1 \\ x_2\end{bmatrix}.
\end{align}
Next a damping value $b$ is determined to maintain a constant damping ratio of $\zeta = 0.01$. 

\begin{figure}[t!]
\centering
\subfloat{
\includegraphics[trim={0mm 0mm 0mm 2mm},clip,width =0.45\columnwidth, keepaspectratio=true]{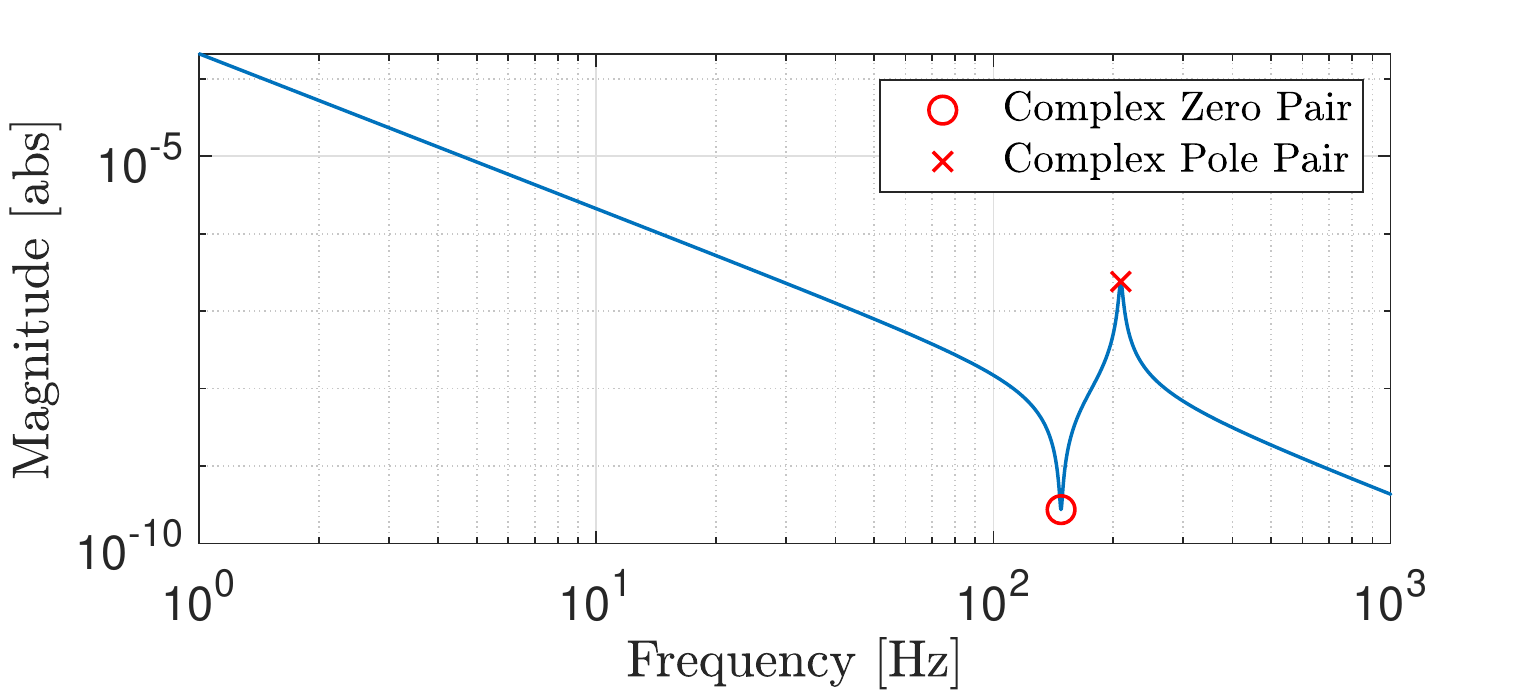}}
\vspace{-4mm}
\caption{Bode plot of two-mass-spring-damping model with $m_1 = m_2 = 60~\rm{kg}$.}
\label{fig:mass_plant}
\vspace{-3mm}
	\centering
	\subfloat[]{\includegraphics[trim={44mm 80mm 55mm 15mm},clip, width =0.45\columnwidth, keepaspectratio=true]{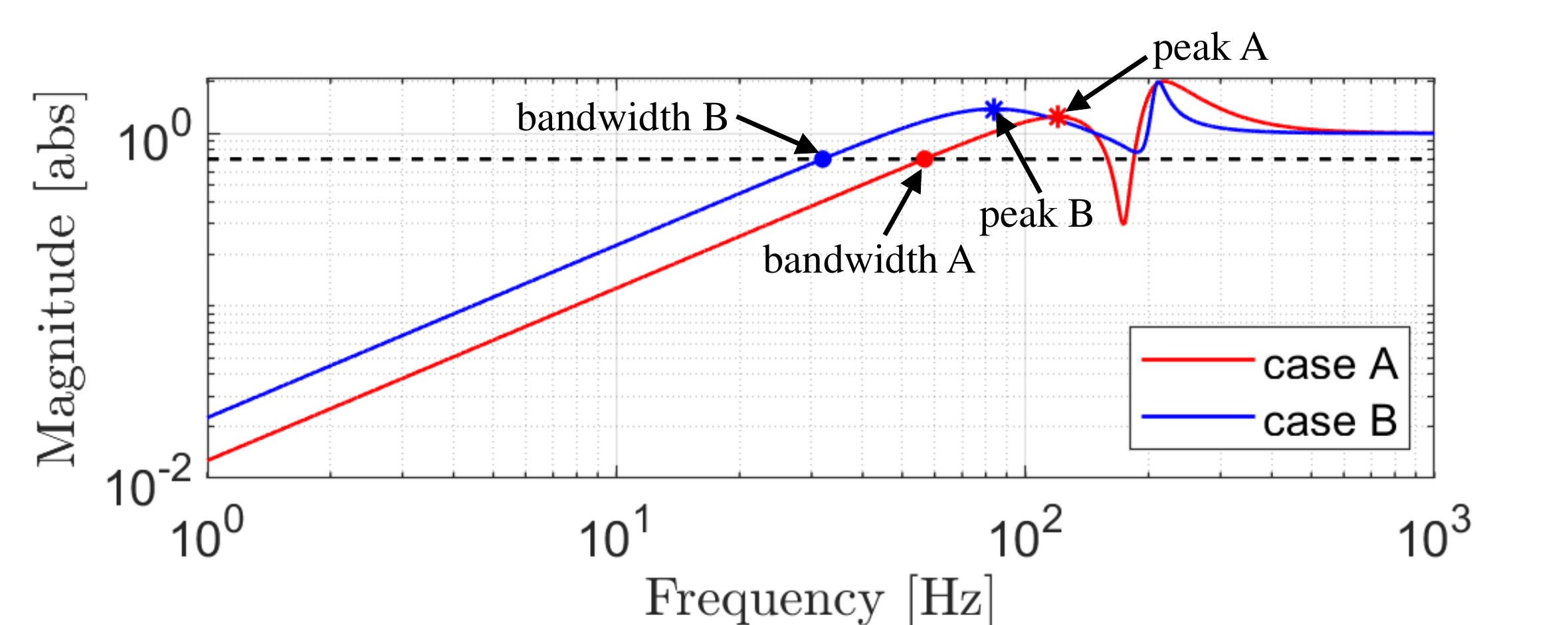}}\\ 
	\vspace{-3mm}
	\subfloat[]{\includegraphics[trim={0mm 0mm 0mm 0mm},clip, width =0.45\columnwidth, keepaspectratio=true]{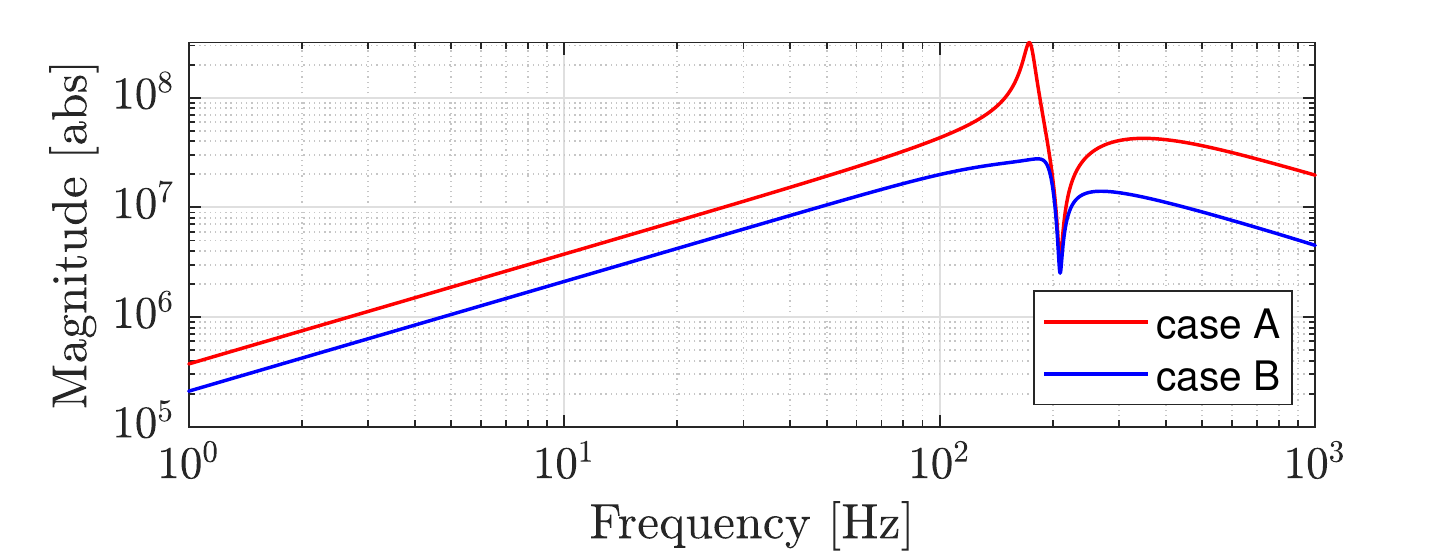}}
	\vspace{-3mm}
	\caption{Comparison between two controllers synthesized with filter parameter $M_{K1} = 2 \times 10^7$ in case~A and  $M_{K2} = 0.5 \times 10^7$ in case B. (a) Closed-loop sensitivity functions. (b) Controller gains. The bandwidth A is at 57~\rm{Hz}, with bandwidth B at 32~\rm{Hz}, peak A at 120~\rm{Hz} and peak B at 84~\rm{Hz}. }
	\label{fig:zero_effect}
	\vspace{-3mm}
\end{figure}

First the dynamics of this system with fixed plant parameters $m_1=m_2=60 ~\rm{kg}$ is considered to identify the limiting factors for control bandwidth, which provides valuable insights for the formulation of the CCD problem. Fig.~\ref{fig:mass_plant} shows the plant frequency response $X_1/F_1$. It can be observed that there exists an anti-resonance at $147~\rm{Hz}$ and a resonance at frequency $209~\rm{Hz}$ in the plant frequency responses, which are caused by the complex zero- and pole-pairs in the system dynamics due to the collocated system configuration. 
Two different controllers are synthesised by solving   \eqref{eqn:inner} with different control effort bound imposed by the weighting filter parameter $M_{K}$, where case~A has $M_{K1} = 2\times 10^{7}$ and case~B has $M_{K2}=0.5\times 10^{7}$. Other weighting filter parameters are fixed  and their values are shown in Table.~\ref{table:filter_para}. 

\begin{table}[t!]
\begin{center}
\begin{footnotesize}
\caption{Mixed sensitivity $H_\infty$ synthesis filter parameters.}
\label{table:filter_para}
\vspace{-4pt}
\begin{tabular}{ |m{0.5cm}| m{0.4cm} | m{0.5cm} | m{0.4cm}| m{0.7cm} | m{0.4cm} |m{0.5cm} |} 
  \hline
   & $M_S$ &  $A_S$   & $c_K$ & $\omega_K$ & $A_l$ & $A_u$\\ 
  \hline
   value & $2$   &  $10^{-4}$   & $10^{4}$ &  $100~\mathrm{Hz}$ & $1$ & $10^{-2}$\\ 
  \hline
\end{tabular}
\end{footnotesize}
\vspace{-2mm}
\end{center}
\end{table} 

Fig.~\ref{fig:zero_effect} illustrates the resultant sensitivity and controller gains of the two controllers. Comparing Fig.~\ref{fig:mass_plant} and Fig.~\ref{fig:zero_effect}, it can be observed that the complex zero pair of the plant causes a peak in the closed-loop sensitivity function in both case~A and case~B, which limits the achievable control bandwidth. This effect is referred to as  the ``limiting effect of the transmission zero''  in the literature \citep{verhoeven2009robust}. Comparing case~A and case~B in Fig.~\ref{fig:zero_effect}, it can be seen that the sensitivity peak caused by the plant complex zeros can be shifted to a higher frequency by a  higher controller gain, thereby achieving  a higher control bandwidth. Note that the resultant control bandwidth still cannot reach beyond the resonance frequency of the complex zeros. This discussion shows that the complex zero frequency is the limiting factor for the achievable control bandwidth in motion systems. 

\begin{rem}
Today's design and control technique for precision motion stages typically takes a sequential approach to design and hardware and controller, and a rule of thumb of ``first structural resonance frequency is 3 to 5 times higher than the target bandwidth'' is typically used \citep{zhou2020magnetically}. Such a sequential approach only considers the frequency of the complex pole pair in the plant dynamics, however, fails to consider the complex zero's effect on control bandwidth. As a result, conservative designs are typically made. 
The CCD methodology provides a promising tool to systemically design the hardware and push for higher control performance. 
\end{rem}

\begin{table}[t]
\begin{center}
\begin{footnotesize}
\caption{Case~Study~\#1 CCD performance evaluation}\label{table:Mass_results}
\begin{tabular}{ | m{3em} | m{1cm}| m{1cm} | m{1.4cm} |m{0.9cm} | m{1.2cm} |} 
  \hline
     & Optimal $m_1$ & Optimal $m_2$  & Bandwidth & Obj. value & 1st res. freq.\\ 
  \hline
  Baseline& 67.56~kg & 67.56~kg& 54.1~Hz & -325.1 & 250.0~Hz  \\ 
  \hline
  Nested CCD & 58.33~kg  & 55~kg & 71.8~Hz & -437.6 & 192.6~Hz \\ 
  \hline
\end{tabular}
\end{footnotesize}
\vspace{-4mm}
\end{center}
\end{table}
The proposed CCD framework in Fig.~\ref{fig:nest_algorithm} is used to determine the controller and hardware designs, where the plant parameters are $m_1$ and $m_2$. 
The hardware objective is selected to be the total weight of the two moving masses as $f(m_1,m_2)=m_1+m_2$. The weights in the overall objective are $w_1 = 0.0995$ and $w_2 = -0.9950$. For inner loop mixed sensitivity $H_{\infty}$ controller synthesis, filter parameters shown in Table.~\ref{table:filter_para} are used, and $M_K = 2 \times {10^7}$. The upper bound for both masses is $70~\rm{kg}$ and lower bound is $55~\rm{kg}$.

To evaluate the performance of the proposed approach, the performance of the CCD approach is compared with that of a baseline design uses a sequential approach where first the hardware is designed and followed by the $H_\infty$ controller synthesis using \eqref{eqn:inner}. 
The baseline target control bandwidth is selected to be $50~\rm{Hz}$, and thus the hardware design optimization aims to have the system's resonance frequency at $250~\rm{Hz}$ according to the rule-of-thumb in sequential design approach, which gives $m_1=m_2=67.56~\rm{kg}$.

Table \ref{table:Mass_results}  shows the comparison result. Since the control coupling and zero effect are not considered in the sequential case, two optimal masses values are large and equal, thus leading to a lower bandwidth. However, in CCD case, $m_2$ is approaching to the lower bound $55~\rm{kg}$ and $m_1$ is converging to $58.33~\rm{kg}$, which shows the different roles of the two parameters. Additionally, the mass values in the CCD case are much lower and the bandwidth is higher since the CCD removes the conservatism and improves the desired performance. 

\subsection{Case Study \#2: Lightweight Motion Stage }

\begin{figure}[t!]
\centering
\subfloat{
\includegraphics[trim={40mm 72mm 40mm 60mm},clip, width =0.5\columnwidth, keepaspectratio=true]{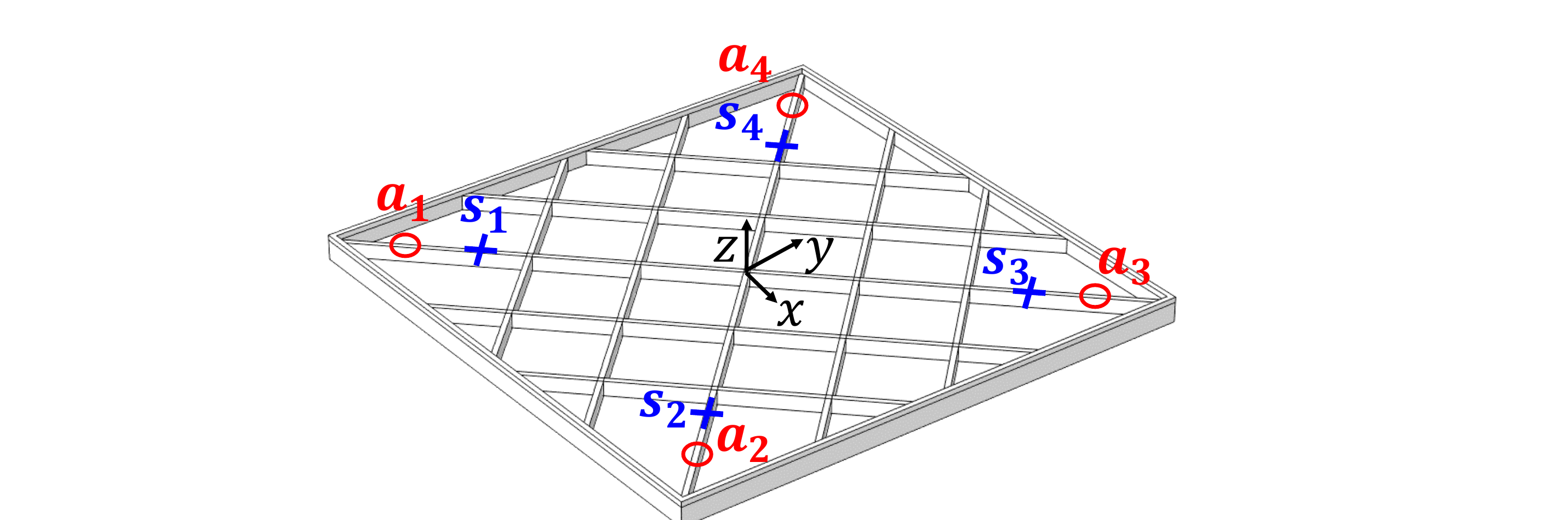}}
\caption{Diagram for actuator and sensor configuration for the rib-enhanced motion stage.}
\label{fig:rib_stage_diagram}

	\centering
	  \captionsetup[subfigure]{justification=centering}
	\subfloat[]{\includegraphics[trim={120mm 80mm 120mm 80mm},clip, width =0.225\columnwidth, keepaspectratio=true]{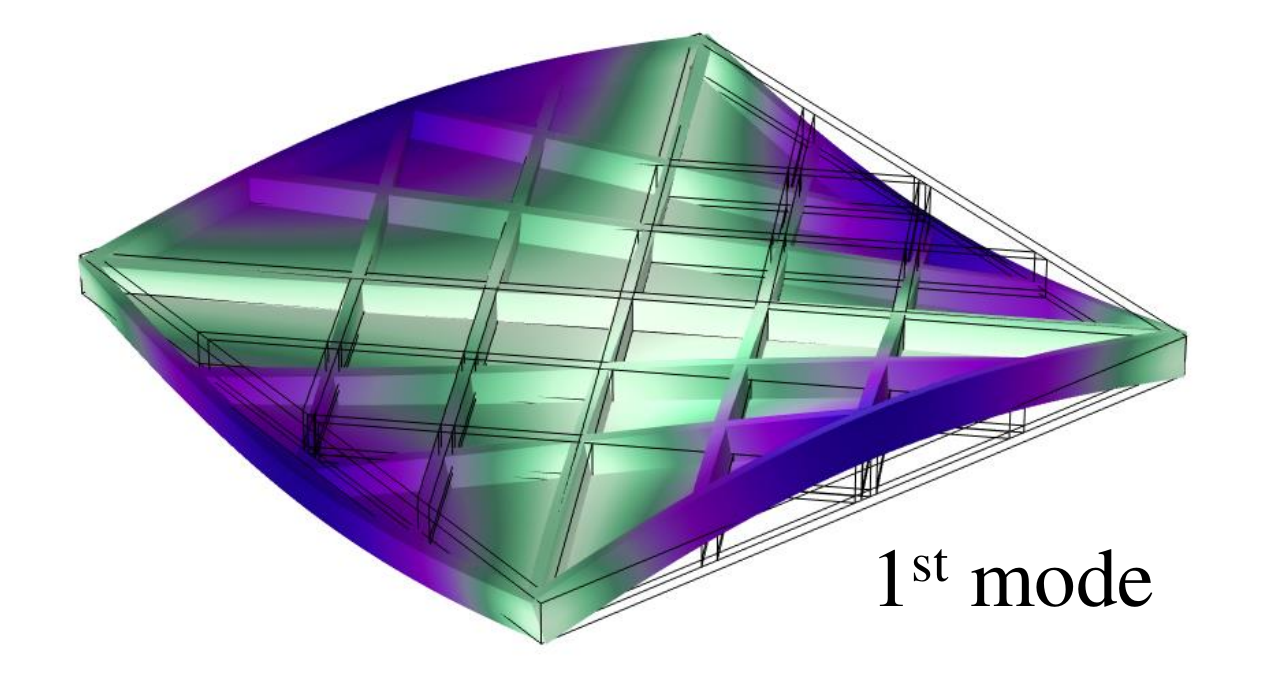}}  
	\subfloat[]{\includegraphics[trim={120mm 80mm 120mm 80mm},clip, width =0.225\columnwidth, keepaspectratio=true]{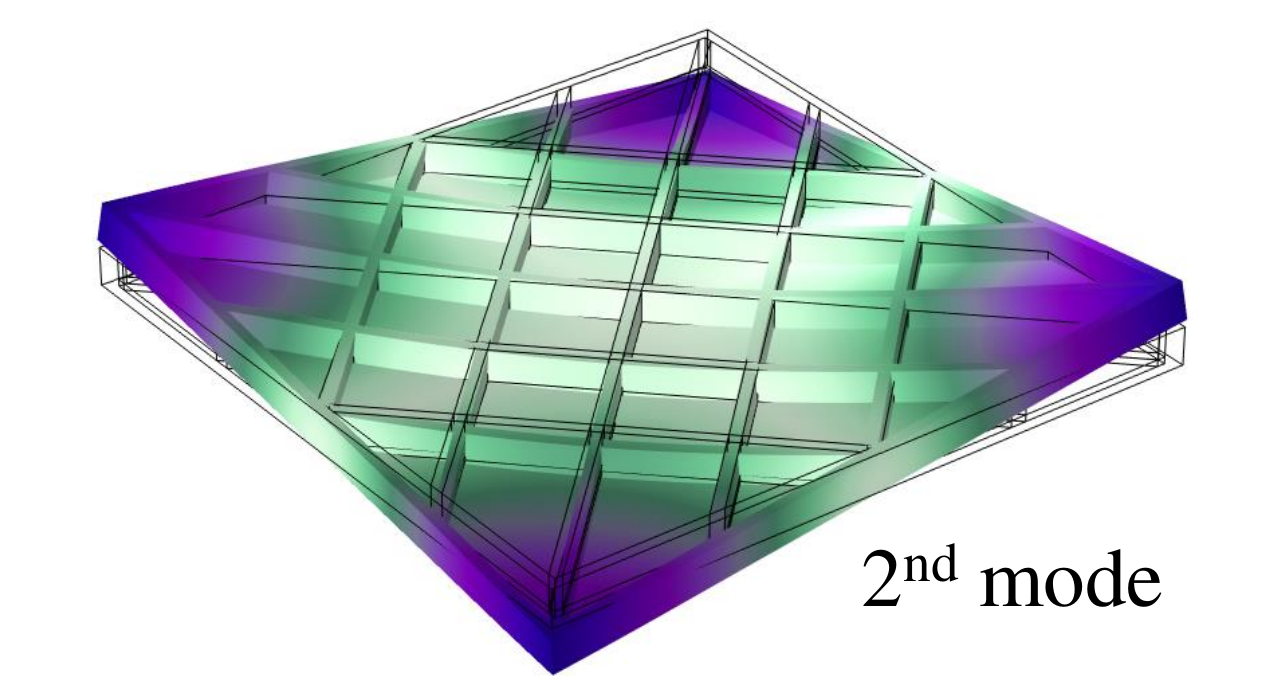}} \\
	\vspace{-4mm}
	\subfloat[]{\includegraphics[trim={120mm 80mm 120mm 80mm},clip, width =0.225\columnwidth, keepaspectratio=true]{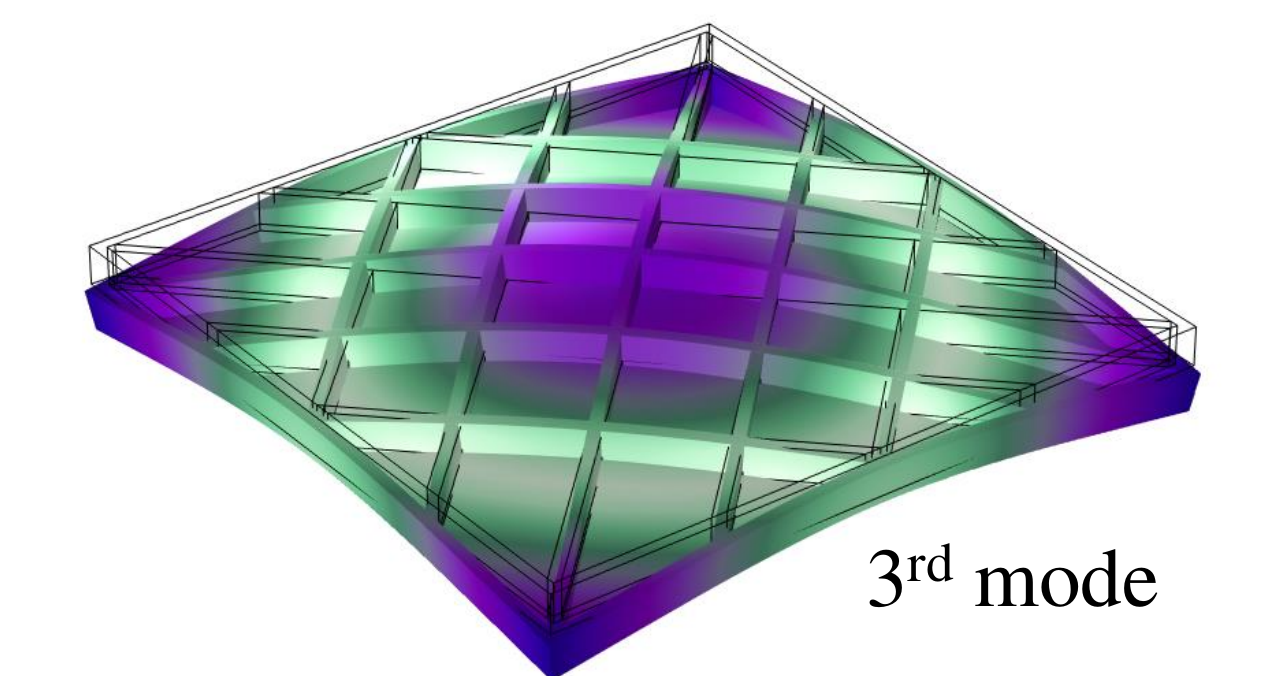}} 
	\subfloat[]{\includegraphics[trim={120mm 80mm 120mm 80mm},clip, width =0.225\columnwidth, keepaspectratio=true]{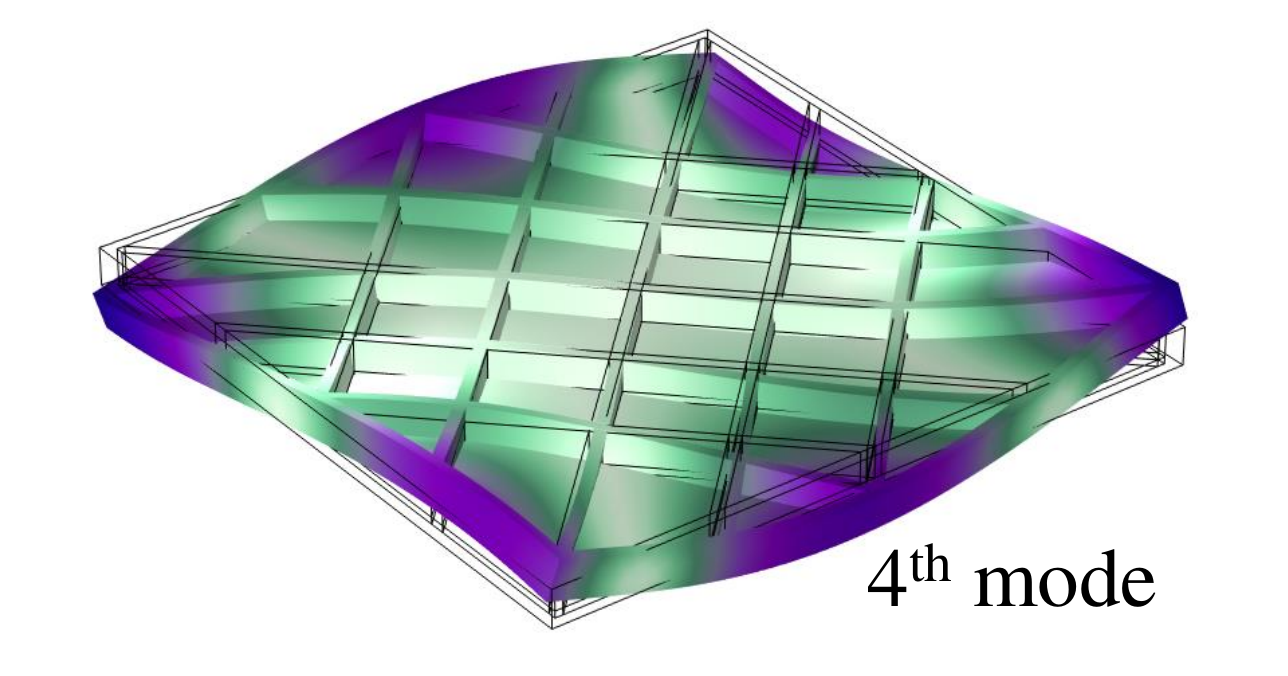}}
	\caption{First four resonance modes of rib-enhanced motion stage. }
	\label{fig:mode_shapes}
	\vspace{-6mm}
\end{figure}

Case study \#2 considers the control and design for a magnetically-levitated precision motion stage as illustrated in Fig.~\ref{fig:rib_stage_diagram}. The stage is made of 6061-T6 aluminum alloy, and the structure is reinforced via ribs. The motion of the stage in three degrees of freedom, including vertical motion, tip, and tilt  (or $x$-, $y$-, and $\theta_z$-directions),  are controlled actively. Four actuators are used to generate the controlling forces and the stage's vertical-directional displacement at four sensors. The position of the actuators and sensors is shown in Fig.~\ref{fig:rib_stage_diagram}. 

Finite element (FE) simulations are  used to obtain the dynamics of the system shown in Fig.~\ref{fig:rib_stage_diagram} to fully consider the stage's spatial-temporal flexible dynamics. 
Fig.~\ref{fig:mode_shapes} shows the first four resonance mode shapes obtained by FE simulations using COMSOL Multiphysics. 
System's undamped equation of motion can be written as
\begin{align} 
  M_{FE}  \ddot{x_{FE}}+K_{FE} x_{FE} &= B_{FE}  u, \label{eq:FE_ODE1}\\
  y &= C_{FE} x_{FE},\label{eq:FE_ODE2}
\end{align}
where $x_{FE}\in\mathbb{R}^{n}$ is a vector for the state variables representing the displacement of the nodes,  $n$ is the number of nodes determined by the FE mesh setting, $M_{FE}$, $K_{FE} \in\mathbb{R}^{n\times n} $ are the mass and stiffness matrices, respectively, which are computed via FE simulations.   $B_{FE}$ and $C_{FE}$ are input and output matrices distributing actuator forces $u$ and measurements $y$ over nodes. 

Note that the system dynamics \eqref{eq:FE_ODE1} has a dimension of $n$. When using a fine mesh in the FE simulations, the value of $n$ is typically very large, and thus directly solving \eqref{eq:FE_ODE2} is computationally expensive. To address this issue, we transform the system dynamics \eqref{eq:FE_ODE1} and \eqref{eq:FE_ODE2} into decoupled modal coordinates as
\begin{align} 
    M\ddot{q}+Kq = Bu, \label{eqn:modal_eom1}\\
  y = Cq,\label{eqn:modal_eom2}
\end{align}
where $q = {\Phi}^{-1}x_{FE}$ is the decoupled modal-coordinate state vector, $\Phi=[\phi_1, \cdots,\phi_n]$ is an $n\times n$ matrix where each column $\phi_i$ represents a vector of the corresponding mode shape, $M =  \Phi^{\top}M_{FE}\Phi$, and $K = \Phi^{\top}K_{FE}\Phi$ are the diagonal model mass and stiffness matrices, respectively,  $B = {\Phi}^\top B_{FE}$ and $C = {\Phi}C_{FE}$ are the decoupled input and measurement matrices, respectively.
With the model coordinates decoupled, we then reduce the system order by truncating the high-frequency flexible modes. In this work, the system dynamics in three rigid-body modes (tip, tilt, and vertical translation) and the first 10 vibration modes are considered, and all  high-frequency modes are truncated. With nominal design parameters, such a model is able to describe the system dynamics accurately up to  1400~Hz, which is sufficient for the control design. Finally, a damping term is introduced to the system such that each flexible mode has a damping ratio of $0.01$. Finally the reduced system dynamic model in form of \eqref{eqn:mech_EOM_1} and \eqref{eqn:mech_EOM_2} can be found.

\begin{figure}[t!]
	\centering
	\vspace{-2mm}
	  \captionsetup[subfigure]{justification=centering}
	\subfloat[]{\includegraphics[trim={120mm 80mm 120mm 45mm},clip, width =0.22\columnwidth, keepaspectratio=true]{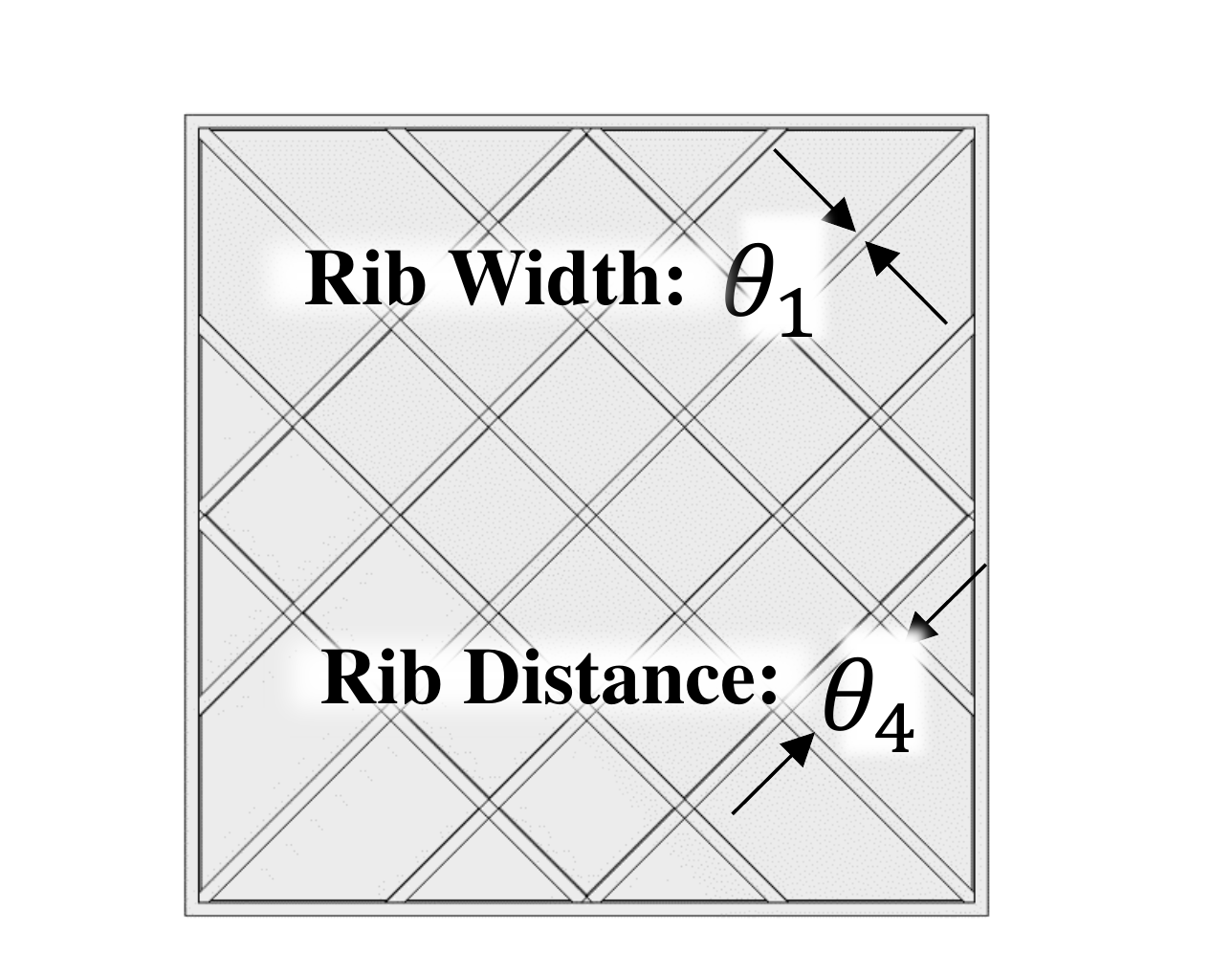}}  
	\subfloat[]{\includegraphics[trim={120mm 60mm 120mm 45mm},clip, width =0.20\columnwidth, keepaspectratio=true]{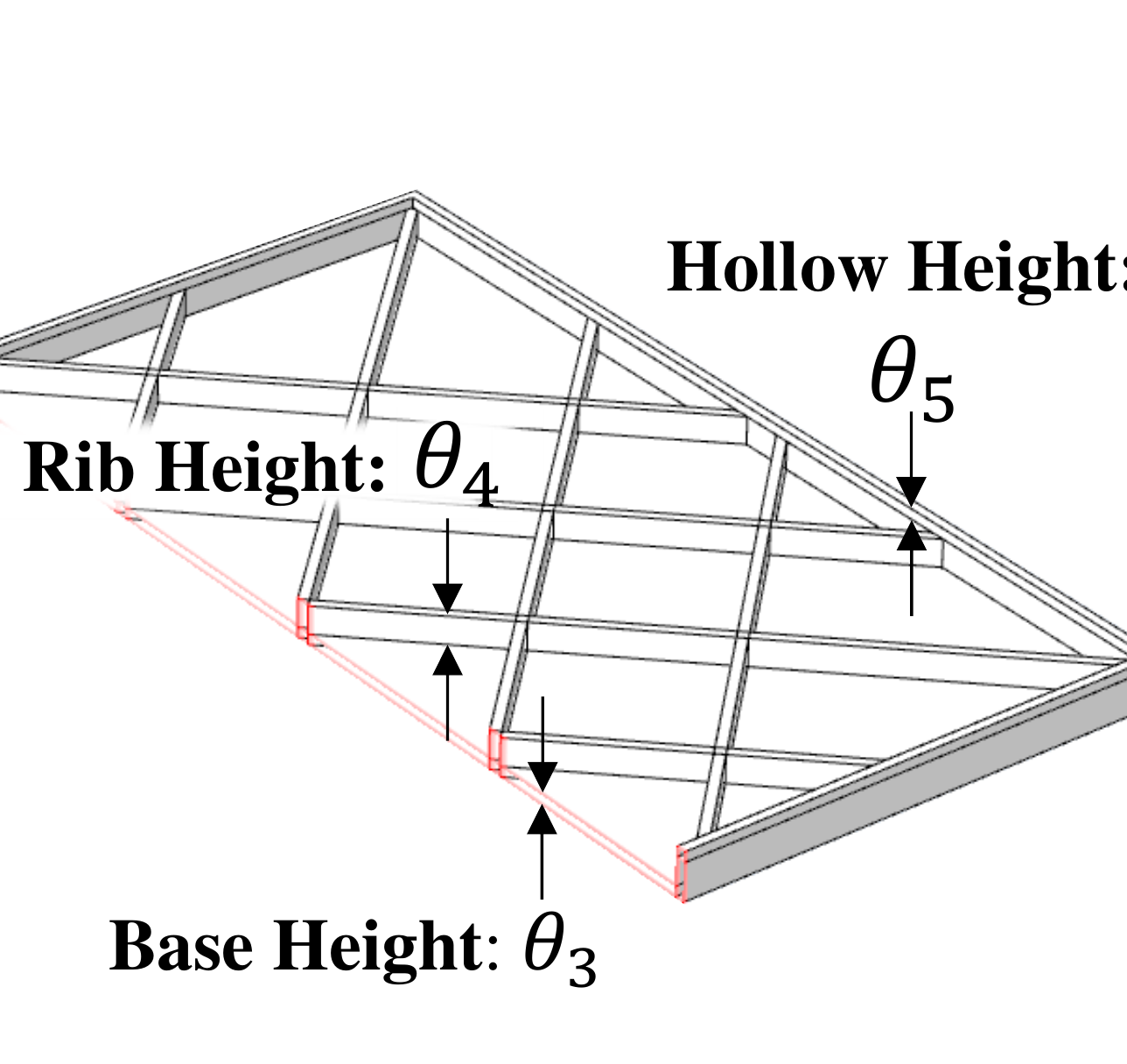}}
	\vspace{-2mm}
	\caption{Hardware design parameters for the rib-reinforced motion stage. (a) Top view (b) Cross-section view. }
	\label{fig:hardware_parameter}
	\vspace{-4mm}
\end{figure}

\begin{figure}[t!]
\centering
\subfloat{
\includegraphics[trim={85mm 60mm 85mm 40mm},clip, width =0.4\columnwidth, keepaspectratio=true]{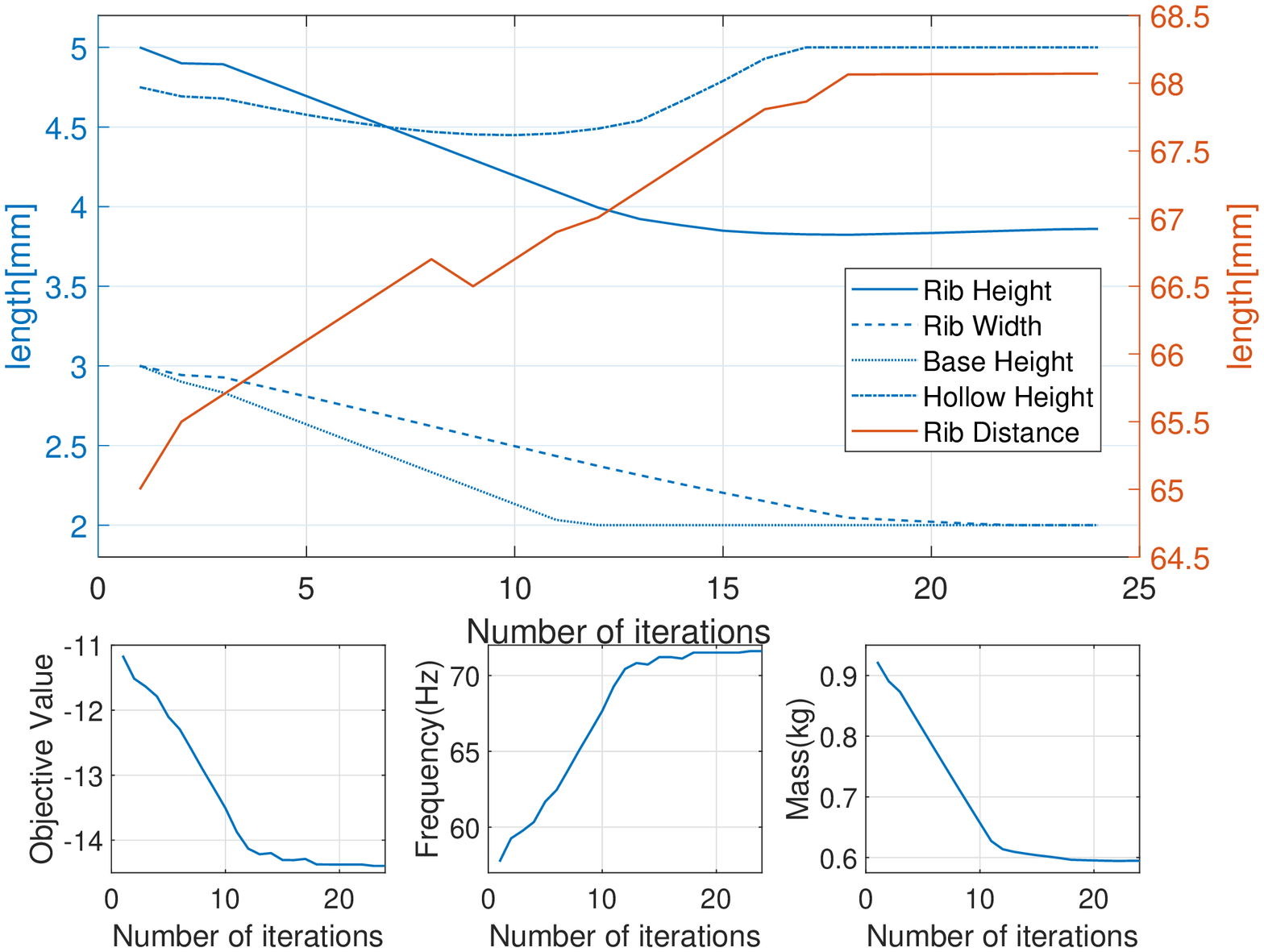}}
\vspace{-4mm}
\caption{History of parameters values during CCD optimization.}
\label{fig:convergence_parameters}


\vspace{-2mm}
\centering
\subfloat{
\includegraphics[trim={5mm 0mm 0mm 5mm},clip, width =0.435\columnwidth, keepaspectratio=true]{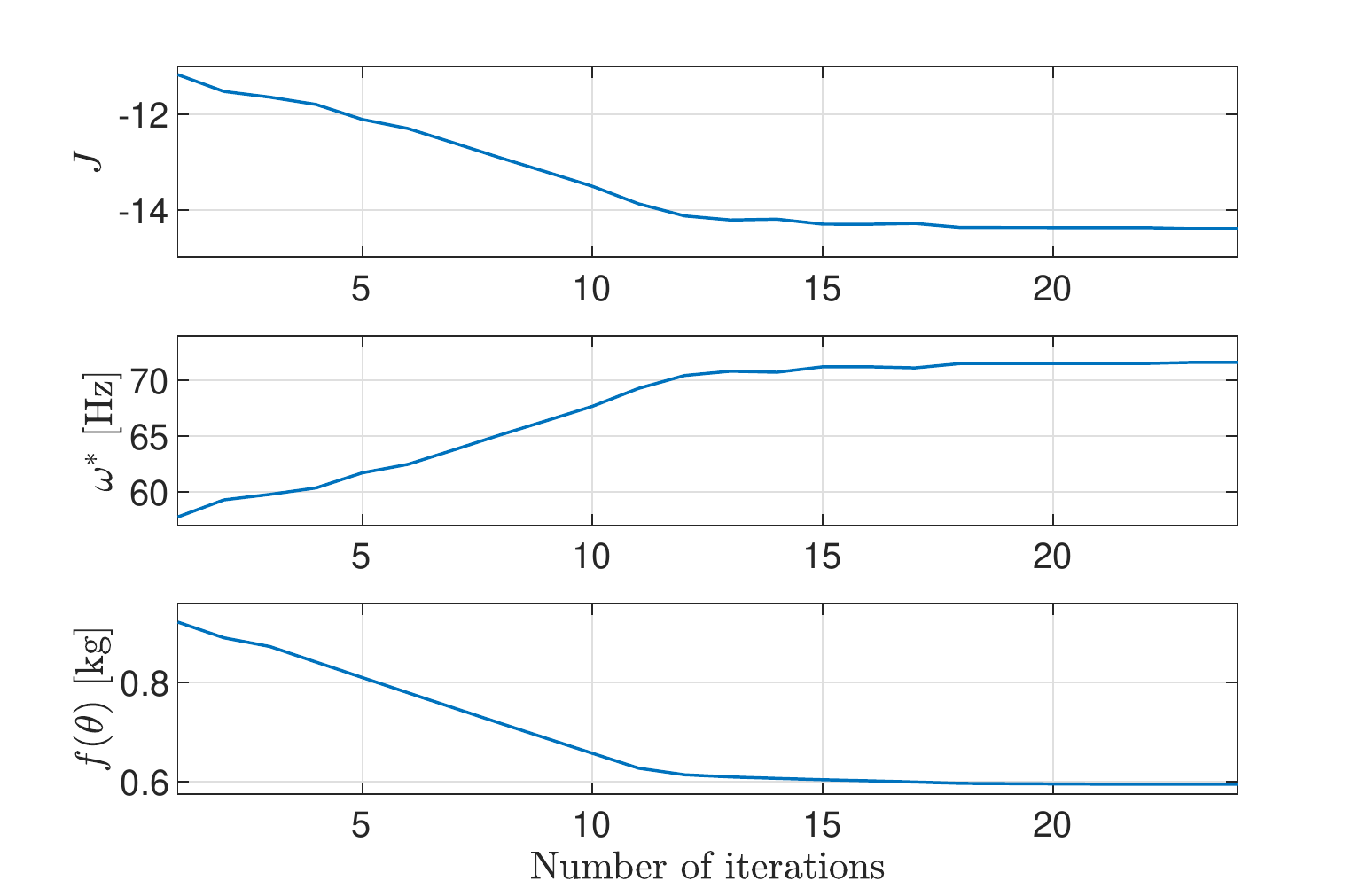}}
\vspace{-4mm}
\caption{History of objective function values during CCD optimization.}
\label{fig:convergence}
\vspace{-2mm}
\end{figure}

The proposed CCD formulation in Fig.~\ref{fig:nest_algorithm} is being used to optimize the motion stage's performance. Here the hardware system cost $f(\theta)$ is chosen to be the total mass of the moving stage,  and the weights between the hardware and control objectives are selected as $w_1 = 0.9994$ and $w_2 = -0.0333$. Fig.~\ref{fig:hardware_parameter} illustrates the selection of hardware parameters $\theta$. 
The inner loop mix-syn $H_{\infty}$ filter values used here are listed in Table.~\ref{table:filter_para} with $M_K = 4.5\times 10^4$. The CCD algorithm took 25 iterations to converge, and Fig.~\ref{fig:convergence_parameters} and \ref{fig:convergence} show the history profile of parameters convergence, objective value, stage's weight, and closed-loop control bandwidth. 

To evaluate the effectiveness of the nested CCD approach, a baseline sequential design is simulated for comparison. In the baseline design, the plant parameters $\theta$ are determined via the shape optimization using the optimization module in COMSOL Multiphysics, where the weight of the stage is being minimized while the stage's first resonance frequency is constrained above 250~Hz. The optimal bandwidth controller is then designed for the given model via solving problem \eqref{eqn:inner}.

\begin{figure}[t!]
\centering
\subfloat{
\includegraphics[trim={95mm 80mm 110mm 60mm},clip, width =0.5\columnwidth, keepaspectratio=true]{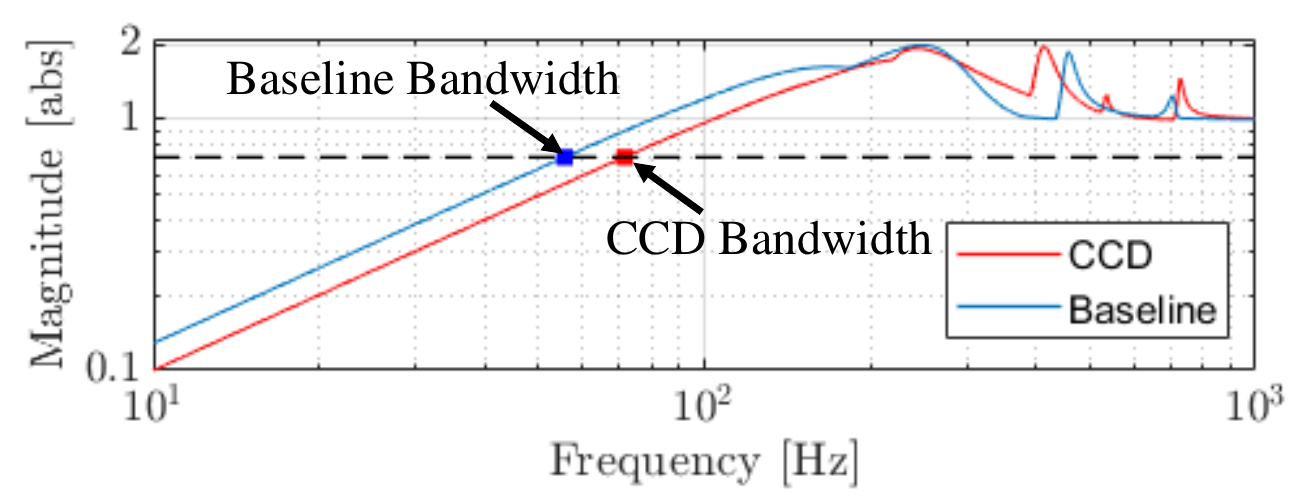}}
\vspace{-4mm}
\caption{Maximum singular value of closed-loop sensitivity functions of the CCD and baseline optimal designs where baseline bandwidth is $55.7~\rm{Hz}$ and CCD bandwidth is $71.6~\rm{Hz}$.}
\vspace{-2mm}
\label{fig:control}
\end{figure}

Fig.~\ref{fig:control} shows a comparison between the maximum singular value for the sensitivity function using the proposed nested CCD design and the sequential design approach, and Table~\ref{table:results} summarizes the performance comparison. It can be observed that the stage weight of the nested CCD case is reduced by 42\% compared to the baseline design, and the closed-loop control bandwidth using the  nested CCD design approach is 28\% higher than that of the baseline case. These results demonstrate that the proposed nested CCD algorithm can successfully optimize the hardware and controller parameters in a unified process, enabling system designs with improved overall performance compared to the conventional sequential design method.

\begin{table}[t]
\begin{center}
\begin{footnotesize}
\caption{Case~Study~\#2  Performance Evaluation.}\label{table:results}
\begin{tabular}{ | m{3.5em} | m{0.8cm}| m{1.2cm} | m{1.2cm} |m{1.3cm} |} 
  \hline
     & Stage Weight  & Bandwidth & Obj. value & 1st res. freq.\\ 
  \hline
  Baseline& 1.02~kg & 55.7~Hz & -10.65 & 250.0~Hz \\ 
  \hline
  Nested CCD & 0.59~kg  & 71.6~Hz & -14.40 & 110.5~Hz\\ 
  \hline
\end{tabular}
\end{footnotesize}
\end{center}
\vspace{-2mm}
\end{table}

\section{Conclusions and Future Work}\label{sec:conclusion}

In this work, we proposed and evaluated a nested CCD framework to achieve desired high-bandwidth control and light moving weight via optimization of the system design and its associated feedback control policy. The proposed framework explicitly optimizes for control bandwidth while incorporating robustness criteria in frequency-domain, which is uniquely suitable for precision motion systems. We also introduced the use of the CCD approach with FE-simulated continuum structural mechanics being considered. We demonstrate the effectiveness of the proposed algorithm with two case studies on motion systems. The results showed that the proposed nested CCD approach can reduce the moving stage's weight by 42\% while improving the control bandwidth by 28\%  comparing with a baseline sequential optimal design approach. Future work will investigate convergence proof of the proposed approach as well as conducting experimental validation for the proposed CCD approach for precision motion systems.

\bibliographystyle{unsrtnat}
\bibliography{references}  






\end{document}